# Bit Error Rate Performance Analysis on Modulation Techniques of Wideband Code Division Multiple Access

M. A. Masud, M. Samsuzzaman, M. A.Rahman

**Abstract -** In the beginning of 21$^{st}$ century there has been a dramatic shift in the market dynamics of telecommunication services. The transmission from base station to mobile or downlink transmission using M-ary Quadrature Amplitude modulation (QAM) and Quadrature phase shift keying (QPSK) modulation schemes are considered in Wideband-Code Division Multiple Access (W-CDMA) system. We have done the performance analysis of these modulation techniques when the system is subjected to Additive White Gaussian Noise (AWGN) and multipath Rayleigh fading are considered in the channel. The research has been performed by using MATLAB 7.6 for simulation and evaluation of Bit Error Rate (BER) and Signal-To-Noise Ratio (SNR) for W-CDMA system models. It is shows that the analysis of Quadrature phases shift key and 16-ary Quadrature Amplitude modulations which are being used in wideband code division multiple access system, Therefore, the system could go for more suitable modulation technique to suit the channel quality, thus we can deliver the optimum and efficient data rate to mobile terminal.

**Index Terms** – Bit Error Rate, Code Division Multiple Access, Quadrature phase shift keying, and Quadrature Amplitude modulation

---

## 1 INTRODUCTION

Wideband Code Division multiple Access (W-CDMA) is being used by Universal Mobile Telecommunication System (UMTS) as platform of the 3rd generation cellular communication system. W-CDMA uses noise-like broadband frequency spectrum where it has high resistance to multipath fading where as this was not present in conventional narrowband signal of 2nd generation (2G) communication system.

High data rate signal transmission can be transmitted over the air by using W-CDMA system, thus enabling of multimedia rich applications such as video streams and high resolution pictures to end users. Thus, we need suitable modulation technique and error correction mechanism to be used in W-CDMA system. In 2G networks, Gaussian Minimum Shift Keying (GMSK) modulation scheme is widely used in GSM (Global System for Mobile) Communication.

This modulation can only transmit data rate of 1 bit per symbol. So it is quite sure that this kind of modulation scheme is not suitable for the next generation communication system. So, there is a need to study the performance of new modulation technique that could deliver higher data rate effectively in a multipath fading channel.

However, the implementation of high data rate modulation techniques that have good bandwidth efficiency in W-CDMA cellular communication requires perfect modulators, demodulators, filter and transmission path that are difficult to achieve in practical radio environment. Modulation schemes which are capable of delivering more bits per symbol are more immune to errors caused by noise and interference in the channel. Moreover, errors can be easily produced as the number of users is increased and the mobile terminal is subjected to mobility. Thus, it has driven many researches into the application of higher order modulations [5]-[10]. Enhance Data Rate for the GSM Evolution (EDGE) is proposed as a transition to 3G as a new Time Division Multiple Access (TDMA) based radio access using the current (800, 900, 1800 and 1900 MHz) frequency bands. EDGE enables significantly higher peak rates and approximately triples the spectral efficiency by

---

- *M. A. Masud is with Dept. of Computer Science & Information Technology, Patuakhali Science and Technology University, Patuakhali, Bangladesh.*
- *M. Samsuzzaman is with Dept. of Computer and Communication Engineering, Patuakhali Science and Technology University,Patuakhali, Bangladesh.*
- *M. A. Rahman is with Faculty of Computer Science & Engineering, Patuakhali Science and Technology University, Patuakhali, Bangladesh.*

employing 8-Phase Shift Keying (8PSK) modulation.

In cellular system, different users have different channel qualities in terms of signal to noise ratio (SNR) due to differences in distance to the base station, fading and interference. Link quality control adapts the data protection according to the channel quality so that an optimal bit rate is obtained for all the channel qualities [5]-[8]. Thus, the system adopts AMC to suit the link quality. W-CDMA systems can employ the high order modulation (8PSK or M-QAM) to increase the transmission data rate with the link quality control.

This research has been focused on the study and the performance measurement of high data rate modulation schemes at those channels which are subjected to Multipath Rayleigh Fading and Additive White Gaussian Noise (AWGN). Modulation Schemes that will be studied are 16-ary QAM (Quadrature Amplitude Modulation) and QPSK (Quadrature Phase Shift Keying). AWGN is the effect of thermal noise generated by thermal motion of electron in all dissipative electrical components i.e. resistors, wires and so on [11]. Mathematically, thermal noise is described by a zero-mean Gaussian random process where the random signal is a sum of Gaussian noise random variable and a dc signal that is

$$Z = a + n \quad \ldots\ldots\ldots\ldots\ldots\ldots\ldots\ldots (1)$$

Where probability distribution function (pdf) for Gaussian noise can be represented as follows where $\sigma^2$ is the variance of $n$.

$$P(z) = 1/\sigma\sqrt{2\pi} \exp[-1/2\{z-a/\sigma\}^2]\ldots\ldots\ldots(2)$$

The effect can cause fluctuations in the received signal's amplitude, phase and angle of arrival, giving rise to terminology multipath fading.

QPSK is one example of M-ray PSK modulation technique (M = 4) where it transmits 2 bits per symbol. The phase carrier takes on one of four equally spaced values, such as 0, $\pi/2$, $\pi$ and $3\pi/2$, where each value of phase corresponds to a unique pair of message bits as it is shown in figure 3.2. The basis signal for QPSK can be expressed as

$$S_{QPSK}(t) = \{\sqrt{E_s}\cos[(i-1)\frac{\pi}{2}]\Phi_1(t) - \sqrt{E_s}\sin[(i-1)\frac{\pi}{2}]\Phi_2(t)\} \quad i=1,2,3,4 \ldots\ldots\ldots\ldots (3)$$

Special characteristics of QPSK are twice data can be sent in the same bandwidth compared to Binary PSK (BPSK) and QPSK has identical bit error probability to that of BPSK. When QPSK is compared to that of BPSK, QPSK provides twice the spectral efficiency with the same energy efficiency. Furthermore, similar to BPSK, QPSK can be differentially encoded to allow non-coherent detection.

QAM is a modulation technique where its amplitude is allowed to very with phase. QAM signaling can be viewed as a combination of Amplitude Shift Keying (ASK) as well as Phase Shift Keying (PSK). The general form of M-ary signal can be defined as

$$S_i(t) = \sqrt{\frac{2E_{min}}{T_s}}a_i\cos(2\pi f_c t) + \sqrt{\frac{2E_{min}}{T_s}}b_i\sin(2\pi f_c t) \quad 0 \leq t \leq T \quad i=1,2,\ldots M \quad (4)$$

Where $E_{min}$ is the energy of the signal with the lowest amplitude integers chosen according to the location of the particular signal point.

A DSSS system spreads the baseband data by directly multiplying the baseband data pulses with a pseudo-noise sequence that is produced by a pseudo-noise (PN) code generator [4]. The PN sequence is usually generated using sequential logic circuits (i.e. feedback shift register). A single pulse or symbol of the PN waveform is called *chip*. Spread spectrum signals are demodulated at receiver through cross-correlation with locally generated version of the pseudo random carrier. Cross-correlation with the correct PN sequence de-spreads the spread spectrum signal and restores the modulated message in the same narrow band as the original data, whereas cross-correlating the signal from an undesired user results in a very small amount of wideband noise at the receiver output.

The performance study will be carried out by varying the chip rate of pseudo noise (PN) generator. W-CDMA (Wideband Code Division Multiple Access) scheme will also be studied by comparing some certain number of users under static and dynamic environment that are subjected to AWGN and multipath Rayleigh fading. The performance of fading channels in W-CDMA system is based on Bit Error Rate (BER) W-CDMA system at downlink transmission and Signal-to-Noise ratio (SNR).

## 2 SYSTEM MODEL

This research has been initiated the reviewing of the high speed data rate on the modulation schemes, DSSS W-CDMA and fading effects on the channels. Then, we have developed a generic model of DSSS W-CDMA as it is shown in fig.1 and is being simulated by MATLAB modulation schemes 16-QAM and QPSK. Both modulation techniques are chosen in this thesis because there are the most important candidates to deliver higher data rate for High Speed Downlink Packet Access (HSDPA), an extension of 3G networks [1]-[3]. The simulation is done under AWGN noise and multipath fading channel using MATLAB 7.6.

As it is shown in fig.1, the user data is assumed to be Bernoulli distributed and can be represented as $b_n(t)$. Each user data is then multiplied with independent or different PN code produced by a PN generator using XOR logical operator. The multiplied signal of each user is represented as $s_n(t)$ after the signal is modulated by either 16-QAM or QPSK. Each signal is added before it is subjected to the channel. At the receiver, the signal $s_k(t)$ is demodulated before the user data is separated from PN code by XOR logical operator. Finally, when the necessary simulations are done, tables and graphs of BER as a function of SNR for various parameters are plotted. Analysis, observations and results will be scaled on plots based on the simulation results. Rayleigh fading and AWGN noise (LOS) are selected to symbolize fading effect in the channel because the system has been done the comparison of W-CDMA system models in two extreme channel conditions.

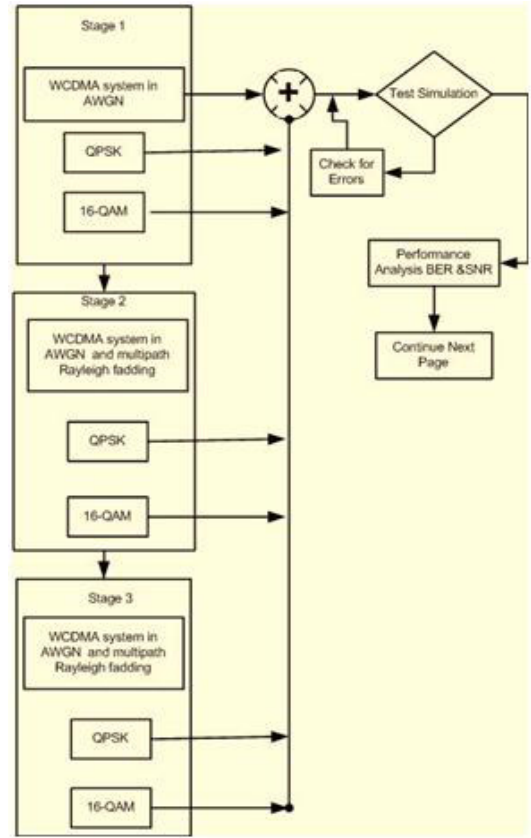

Fig. 1. Simulation flowchart for W-CDMA system models used in Simulink and M files.

There are many fading effects that can be categorized as large-scale and small-scale fading. Rayleigh fading represents the worst case of multipath fading where it represents small-scale fading due to small changes in position with respect to time that is Doppler Effect. On the other hand, AWGN represents the thermal noise generated by electrical instruments.

As computer based simulations are the most fitting, powerful and proficient means to stand for the actual or real time scenarios of mobile radio system. Thus, MATLAB 7.6 has been used to simulate W-CDMA model based on associated parameters, theories and formulae. Throughout this work, we kept the bit rate of 384Kbps for the signal generator. There will be three W-CDMA wireless cellular system models that will be used in this research. The models are W-CDMA system in AWGN channel and W-CDMA system in AWGN and Multipath Rayleigh Fading.

There are many different ways to characterize CDMA codes, but nothing can be more intuitive and effective than the auto-correlation function (ACF) and cross-correlation function (CCF). The ACF is defined as the result of chip-wise convolution, correlation or matched-filtering operation between two time-shifted versions of the same code, which can be further classified into two sub-categories: periodic ACF and aperiodic ACF, depending on the same and different signs of two consecutive bits. Usually the periodic and aperiodic ACFs appear equally likely due to the fact that the binary data of '+1' and '−1' always appear with equal probability in binary bit streams. The cross-correlation function (CCF) is defined as the result of a chip-wise convolution operation between two different spreading codes in a family of codes. For a similar reason to that mentioned earlier for the ACF, there are also two different types of CCF, i.e. periodic and aperiodic CCF. The argument of this function is the number of periods of the code for which the autocorrelation function is to be obtained. The equation-4 is used to calculate the value of cross-correlation function between two distinct codes $X(t)$ and $Y(t)$.

$$R_{xx}(t) = \frac{1}{T}\int_0^T X(t)Y(t+\tau)\,d\tau \ldots\ldots\ldots (5)$$

The arguments of this function are the name of the sequence and the number of periods of the code for which the autocorrelation function is to be obtained. The following function will be typed to calculate the cross-correlation function of codes $X(t)$ and $Y(t)$.

In this task Linear feedback shift register will be used to generate code sequences in WCDMA. A shift register contains a number of cells identified by numbers 1 to r, and each cell is a storage unit that, under the control of a clock pulse, moves the contents to its output while reading its new contents from the input. In a standard configuration of a feedback register, the input of cell $m$ will be a function of the output of cell $m$-1 and the output of cell r (the last cell of the shift register) forms the desired code sequence.

In linear feedback shift registers (linear FSRs), the function combining the outputs of cell m-1 and cell r with the input of cell m is linear. Fig. 2 shows a single linear binary shift register, which can generate a sequence from generation polynomial $h(x) = x5 + x2 + 1$. In general, the configuration of a linear binary shift register of $n$ sections is described by a generator polynomial, which is a binary polynomial of degree $n$. $n$, in this case, is the number of register of the shift register.

$H(x)=h_nx^n+h_{n-1}x^{n-1}+\ldots h_1x^1 +1$ ( $h_i \in \{0,1\}$..(5)

Also, in this case, three-stage M-sequence and a random sequence with a code length of 7 will be used. M-sequence is a sequence generated by a single LFSR where a sequence of possible period, (Nc= $2n$ -1), is generated by an n-stage binary shift register with linear feedback. To generate an M-sequence, the generator polynomial must be of degree $n$. Thus, the periodic autocorrelation function of an M-sequence is given by

The system is configured based on synchronous W_CDMA system. Each user employs their own sequence to spread the information data. In this downlink transmission, the information data are modulated by modulation scheme.

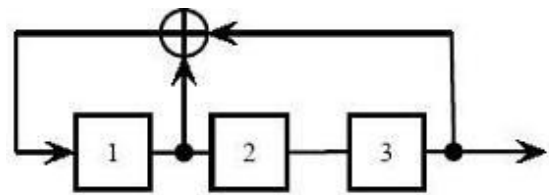

Fig. 2. Three-stage M-sequence

After, the modulated data are spreaded by the code that is M-sequence. The spreaded data of all users in the system are transmitted to the mobile users at the same time. The mobile user detects the information data of each user by correlating the received signal with a code sequence allocated to each user. The performance of W-CDMA system is studied based on QPSK and 16-QAM modulation techniques that has been used in this simulation.

## 4 RESULT AND DISCUSSION

Based on data generated by computer simulation of W-CDMA models, relationship for model using QPSK and QAM modulation techniques between BER as a function of the following parameters are obtained.

TABLE 1
Simulation result for evaluation on BER vs. SNR

| Signal to Noise Ratio(Eb No) | Number of Error | Bit Error Rate(BER) |
|---|---|---|
| 0 | 15615 | 7.807500e-002 |
| 1 | 11334 | 5.667000e-002 |
| 2 | 7520 | 3.760000e-002 |
| 3 | 4481 | 2.242000e-002 |
| 4 | 2489 | 1.444000e-002 |

The simulation is followed by using m file. In this approach, the simulation is successfully done using QPSK modulation technique. The desired BER graphs are obtained for simulation in AWGN channel. QPSK and 16-QAM modulation techniques in AWGN channel has good performance when it is compared to that of Multipath Rayleigh channel. Also, the performance of QPSK and 16-QAM degrades when the channel is subjected to Multipath fading with increasing value of Doppler shift (Hz).

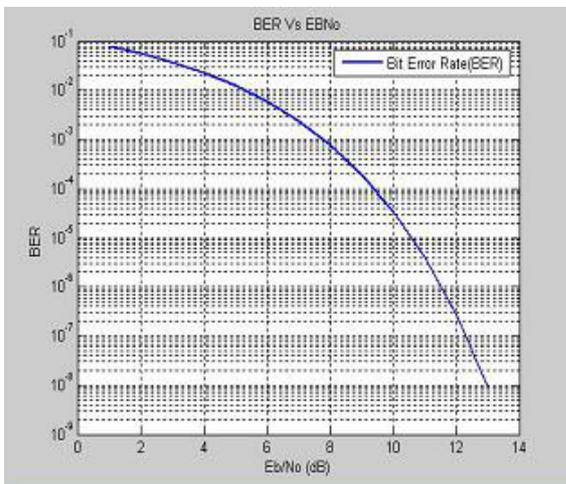

Fig. 3. AWGN Channels in W-CDMA

In other words, it performs poorly as the speed of mobile terminal is increased. Moreover, the system performs badly as the number of users is increased. Comparison between QPSK and 16-QAM modulation schemes shows that 16-QAM performs very poorly in both AWGN (LOS channel) and AWGN with Multipath fading channel.

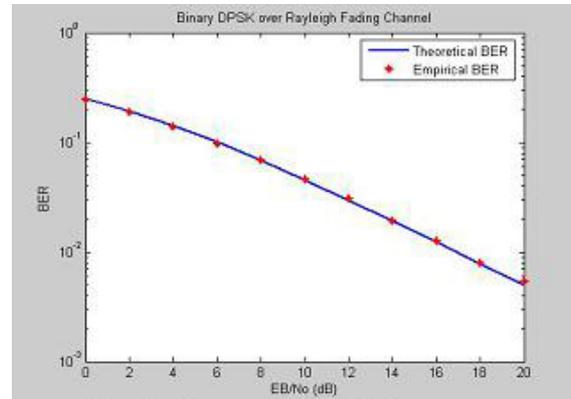

Fig. 4. Performance of W-CDMA in Multipath Rayleigh Fading Channels

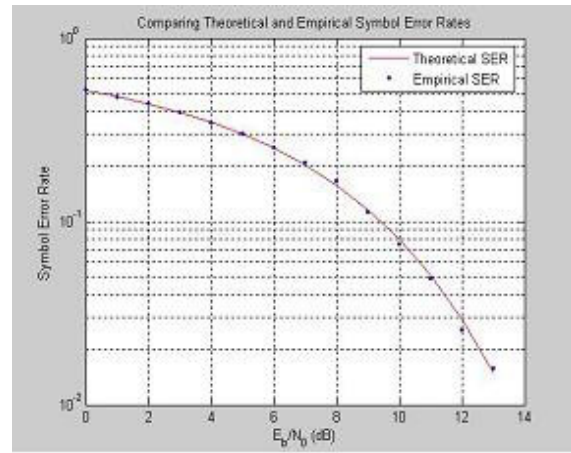

Fig. 5. Comparison of Theoretical and Empirical Symbol Error Rates

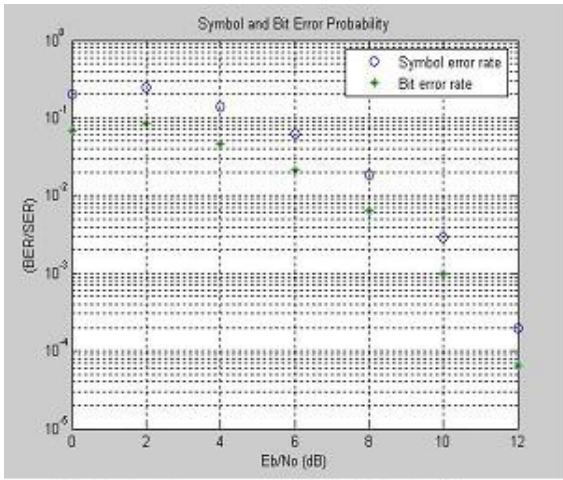

Fig. 6. Performance of BER/SER for 16-QAM with gray coding

The simulation of 16-QAM modulation technique using m files cannot be done because it is suspected that the variation of amplitude with phase causes errors in the constellation of 16-QAM signal. The reason behind this poor performance of 16-QAM of W-CDMA system in multipath fading channel is basically due to the interference between adjacent carriers phase in the constellation of 16-ary QAM.

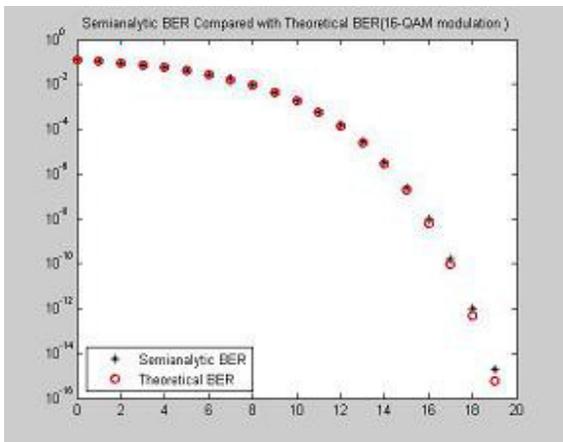

Fig. 7. Comparison of Semianalytic BER and Theoretical BER (16-QAM modulation)

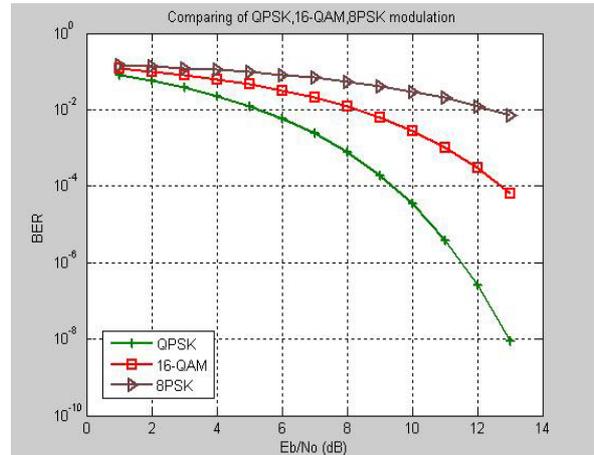

Fig. 8. Performance Comparison of QPSK, 16-QAM and 8PSK Modulation

A sound approach is needed to be used in 16-QAM of W-CDMA system to ensure zero or minimal interference between adjacent carriers phase in the constellation of 16-QAM. It is suggested that error correction coding such as convolution coding or turbo coding is used in this system to ensure better performance of 16-QAM modulation technique of W-CDMA system. Also, it is possible to consider the use of a RAKE receiver or a smart antenna (MIMO) in this system to exploit the delayed signals generated in multipath fading channel. It is discovered, as well, that the performance of multi-user in the m file is limited to a maximum of 7 users. Thus, this system needs to be improved to simulate more number of users so that the performance of multiple accesses in W-CDMA can be studied more dynamically.

## 5 CONCLUSION

In telecommunication field the major challenges is to convey the information as efficiently as possible through limited bandwidth, though the some of information bits are lost in most of the cases and signal which is sent originally will face fading. To reduce the bit error rate the loss of information and signal fading should be minimized. In our thesis we analyze two modulation techniques, QPSK and 16-QAM to reduce the error performance of the signal and compare which technique is better through Rayleigh Fading Channel in the presence of AWGN. The performance of W-CDMA system in AWGN channel shows that QPSK modulation technique has a better performance compared to that of 16-QAM. Furthermore, similar trend is found when the channel is subjected to multipath Rayleigh fading with Doppler shift. The performance of QPSK and 16-QAM modulation technique in W-CDMA system degrades as the mobility is increased from 60kmph to 120kmph for both QPSK and 16-QAM. However, QPSK shows better performance compared to that of 16-QAM in LOS channel and multipath Rayleigh fading channel.

In other words, 16-QAM suffers signal degradation and error probed when the simulations are done in these channels. As the number of users is increased, the QPSK modulation technique performs poorly in W-CDMA system. Unfortunately, the simulation for 16-QAM has failed to show the expected results in both Simulink and m files. This is because the 16-QAM modulation scheme experiences adjacent carrier interference when the simulation is carried out. Therefore, it results in inconsistence of data or signal throughput causing abnormal values of BER and eventually affecting the performance of W-CDMA system. It is expected that 16-QAM will show performance degradation similar like QPSK as the number of users is increased but with lower performance compared to that of QPSK.A more complete W-CDMA system can be developed using the suggested method as they are explained as follows.

- Generate binary data source for various data rates for various services that can be offered by W-CDMA system in 3G environment.
- Implement error correction scheme such as convolution coding and turbo coding Particularly with M-QAM modulation technique in W-CDMA system. Higher order QAM modulation schemes are vulnerable to error. Therefore, error correction Coding ensures higher chances of signal survivability in AWGN and multipath Rayleigh Channel and thus enhances the performance of the system.
- It is proposed that Rician fading is included in the channel in addition of AWGN and Multipath Rayleigh fading channel. Then, comparison can be made between these channels.

Block set are many, more designs of block set using CDMA, especially W-CDMA technologies are needed in this research. This is to produce high accuracy and precision simulation model of W-CDMA system.

## ACKNOWLEDGEMENT

We would like to cordial thanks to dean, faculty of Computer Science and Engineering, chairman, dept. of Computer Science and Information Technology. The research work supported by communication laboratory and its annual allowance of Patuakhali Science and Technology University, Bangladesh.

## REFERENCES

[1] J. M. Holtzman, "A Simple, Accurate Method to Calculate Spread-Spectrum Multiple- Access Error Probabilities," IEEE Trans. Communication, vol. 40, pp. 461- 464, Mar.1992.

[2] Victor Wen-Kai Cheng, Wayne E. Stark, "Adaptive Coding and Modulation for Spread Spectrum", IEEE Journal, 1997.

[3] Troels E. Kolding, Frank Frederiksen, Preben E. Mogensen, "Performance Aspects of WCDMA Systems with High Speed Downlink Packet Access (HSDPA)", Nokia Network R&D, Denmark, 2003

[4] Crovella, M.E., Bestavros, A., "Self-Similarity in World Wide Web Traffic: Evidence and Possible Causes", IEEE/ACM Trans. on Networking, Vol. 5, No. 6, pp. 835-846, December, 1997.

[5] Y.Rosmansyah, P. Sweeney, R. Tafazolli, "Air-Interface Techniques for Achieving High Data Rates for UMTS", IEEE 3G Mobile Communication Technologies, Conference Publication No. 477, pp. 368-372, 26-28 March 2001.

[6] A.S. Madhukumar, Francois Chin, "An Efficient Method for High-rate Data Transmission using Residue Number System based DS-CDMA", IEEE.

[7] Min-yan Song, Yang Xiao, Joachim Habermann, "High Data Rate Wireless System", IEEE, pp. 1344-1350.

[8] Haifeng Wang, Zhenhong Li, "Novel Soft-bit Demodulator with Multi-dimensional Projection for High-order Modulation", IEEE, pp. 2051-2054, 2002.

[9] Troels Emil Kolding, Klaus ingemann Pedersen, Jeroen Wigard, Frank Frederiksen, Preben Elgaard. Mogensen, "High Speed Downlink Packet Access (HSDPA): W-CDMA Evolution", IEEE Vehicular Technology Society News, February, 2003.

[10] E. Hossain, T. Issariyakul, "Performance bound of dynamic forward link adaptation in cellular W-CDMA networks using high-order modulation and multicode formats", IEEE Electronics Letters, Vol.40, No. 2, January 2004.

[11] Bernard Sklar, "Digital Communications: Fundamentals and Applications", Prentice- Hall, 2nd Edition, pp. 30-33.

**M. A. Masud** received the B.Sc. and M.Sc. degree in Information and Communication Engineering from Islamic University, Kushtia, Bangladesh in 2005 and 2006 respectively. Since March 2007, he has been on the faculty of the Department of Computer Science and Information Technology at Patuakhali Science and Technology University, Bangladesh. His current research interests are Cellular Communication, WCDMA, and OFDMA technology.

M. Samsuzzaman completed the B.Sc. and M.Sc. degree in Computer Science and Engineering from Islamic University, Kushtia, Bangladesh in 2007 and 2008 respectively. He has been working as a lecturer in the department of Computer and Communication Engineering at Patuakhali Science and Technology University since 0n Feb. 2008. His research interests are WSN, WCDMA and GSM.

**M. A. Rahman** Completed B. Sc. in Computer Science and Engineering from Patuakhali Science and Technology University.